\documentclass[%
reprint,
showpacs,
nofootinbib,
 amsmath,amssymb,
 aps,
]{revtex4-1}
\usepackage[colorlinks=true,linkcolor=blue]{hyperref}
\usepackage{graphicx}
\usepackage{dcolumn}
\usepackage{bm}
\usepackage{subfigure}

\begin{document}


\title{Fermion Resonances on a Thick Brane with a Piecewise Warp Factor}


\author{Hai-Tao Li}
\email{liht07@lzu.cn}
\author{Yu-Xiao Liu}%
\email{liuyx@lzu.edu.cn}
\thanks{corresponding author.}
\author{Zhen-Hua Zhao}
\email{zhaozhh02@gmail.com}
\author{Heng Guo}
\email{guoh06@lzu.cn}
\affiliation{%
Institute of Theoretical Physics, Lanzhou University, Lanzhou 730000, People's Republic of China
}%

\date{\today}

\begin{abstract}
In this paper, we mainly investigate the problems of resonances of
massive KK fermions on a single scalar constructed thick brane with a piecewise warp factor matching smoothly. The distance between two boundaries and the other parameters are determined by one free parameter through three junction conditions. For the generalized Yukawa coupling $\eta\overline{\Psi}\phi^{k}\Psi$ with odd $k=1,3,5,\ldots$, the mass eigenvalue $m$, width $\Gamma$,
lifetime $\tau$, and maximal probability $P_{max}$ of fermion
resonances are obtained. Our numerical calculations show that the brane without internal structure also favors the appearance of resonant states for both left- and right-handed fermions. The scalar-fermion coupling and the thickness of the brane influence the resonant behaviors of the massive KK fermions.
\end{abstract}

\maketitle

\section{Introduction}

As we know, in order to unify electromagnetism with Einstein gravity, Kaluza and
Klein first proposed that space-times have more than three spatial
dimensions~\cite{Kaluza}. Later, higher dimensional space-time with large extra
dimensions~\cite{Akama:82a,Rubakov:83a,Rubakov:83b,Akama:83a,Visser:85a,
Randjbar:86a,Antoniadis:90a,Arkani-Hamed:98a,Kokorelis,Randall:99a,Randall:99b,
Arkani-Hamed:00a,Lykken:00a,Kehagias:04a,Mannheim} had been paid more attention; furthermore,
the emphasis shifted to a ``brane world''
picture
which opened up a rich and interesting route towards solving the long-standing problems
like the mass hierarchy and cosmological constant problems in high-energy physics.
Subsequently, the idea had been applied in the standard model~\cite{Akama:82a,Rubakov:83a,Kokorelis}. In particular, inspired also
in~\cite{Horava:96a,Horava:96b}, it was put forward that at the energy scales of
standard model the matter fields cannot propagate into extra dimensions whilst
gravity on the other hand can permeate through all dimensions~\cite{Arkani-Hamed:98a,Randall:99a,Randall:99b}.
A new wave of
research in the field of extra dimensions came with the framework of Arkani-Hamed,
Dimopoulos, and Dvali who proposed the large extra dimensions model (ADD
model)~\cite{Arkani-Hamed:98a}, which lowers the energy scale of quantum gravity to
1 TeV by localizing the standard model fields to a 4-brane in a higher dimensional
space-time so that the hierarchy problem can be addressed within the framework of
ADD model. In Ref.~\cite{Randall:99a}, the Randall-Sundrum (RS) model provides an
alternative explanation for the gauge hierarchy problem in particle physics based on
small extra dimensions and a nonfactorizable bulk geometry ($\text{AdS}_{5}$). Subsequent
progress about brane world and extra dimension suggested that the warped metric
could even provide an alternative to compactification for the extra
dimensions~\cite{Randall:99b, Arkani-Hamed:00a}.

In RS warped brane world scenarios, the brane is the so-called thin brane, which is
an infinitely thin object, in which the energy density of the brane is a deltalike
function with respect to a fifth dimension coordinate. So this model is a very idealized
brane world model. Recently, more realistic thick brane models were investigated
in higher dimensional space-time
\cite{DeWolfe:00a,Gremm:00a,Gremm:00b,Csaki:00a,Campos:02a,WangPRD2002,thickBranes,Guerrero:02a,Dzhunushaliev:08a,Bazeia:03a,Shtanov:09a,Shifman:09,Alfredo:09}.
In thick brane scenarios, the coupling between gravity and scalars should be
introduced. By introducing scalar fields in the bulk \cite{GW}, the modulus can be
stabilized in a RS warped brane world scenario. In the thick brane scenario, the known
solutions can be classified into topologically nontrivial solutions and trivial
ones. A comprehensive review on the thick brane solutions and related topics is
given in Ref.~\cite{Dzhunushaliev:09a} and the definition of thick brane was given
to avoid the problems related to possible different understandings of this
term.

Then, it is possible to describe the matter and interactions localized on the brane
by a natural mechanism in higher dimensional space-time. One can consider various
bulk fields on the brane, such as spin $0$ scalars, spin $1/2$ fermions, and so
forth~\cite{Pomarol:00,Ringeval:02,Melfo:06,Slatyer:07a,Liu:08b,Liu:08c,Liu:09a,Zhao:09a,Koley:09}. However, the localization of spin $1/2$ fermions shows very interesting and
important properties. In addition, the localization problem of fermions on a kind
of topological defect called domain wall has been extensively
investigated~\cite{Ringeval:02,Melfo:06,Slatyer:07a,Liu:08b,Liu:08c,Liu:09a,Zhao:09a,Koley:09}. For
localizing fermions on the thick branes or domain walls, the other kind of
interactions were introduced in addition to gravity. The common interactions include
the generalized Yukawa coupling between the fermions and the background scalar
field. Besides, one can consider the gauge field~\cite{Parameswaran:07a,Liu:07a}
localized in our universe. The localization problem can also be discussed in the context of supersymmetry
and supergravity~\cite{Cvetic:1992bf,Cvetic:1996vr,dePol:00a}.
In different backgrounds, such as vortex
background~\cite{Liu:07b,Wang:05a,Rafael:08a,Starkman:02a} and general space-time
background~\cite{Randjbar-Daemi:00a}, the localization of various fields was
investigated.

The discontinuous Kaluza-Klein
(KK) modes with gap called bound states and the continuous gapless states with
$m^2>0$ can be
derived~\cite{Barbosa-Cendejas:08a,Liu:08a,Kodama:09a,20082009}. Besides, the
metastable KK states of the graviton or fermion with finite lifetime can also be
obtained in many brane world models~\cite{Csaki:00a,Ringeval:02, Gregory:00,Clarkson:05,
Davies:07, Bogdanos:09,Bazeia:06a, 0801.0801, Almeida:0901, Liu:09b, Liu:09c, Liu:09d, Cruz:09a, Cruz:09b}. Recently, the problem of the fermion resonances has been extensively
studied in many thick brane models~\cite{Davies:07,Almeida:0901,Liu:09b,Liu:09c,Liu:09d,Cruz:09b,Liu:10a}. In Ref.~\cite{Davies:07}, the authors investigated the general
properties of localization of fermions and scalars in smoothed field-theoretical
versions of the type II RS brane world model. They reached an important conclusion
that, if discrete bound states are present in the gravity-free case, those become
resonant states in the continuum, while off-resonant modes are highly suppressed on
the brane. In Ref.~\cite{Almeida:0901}, the authors investigated fermion resonances
on the Bloch brane constructed with two background scalar fields~\cite{Bazeia:04a}.
The problem of the massive resonant KK modes with even-parity for both chiralities
was studied in depth and their appearance is related to branes with internal structure. In Ref.~\cite{Liu:09b}, the fermion
resonances were studied in the context of a de Sitter thick
brane; the authors found that there exist resonant KK modes in a single brane without internal structure. The two conclusions are different because in Ref.~\cite{Liu:09b} the coupling parameter $\eta$ was varied but in Ref.~\cite{Bazeia:04a} $\eta$ was fixed. Particularly, in Ref.~\cite{Liu:09d}, the
authors extended the fermion resonances to multiscalar generated thick branes and
found the Kaluza-Klein chiral decompositions of massive fermion resonances are the
parity-chirality decompositions. In this paper, we consider the fermion resonances on
the thick brane model with a piecewise warp factor~\cite{0801.0801}. In addition to the conventional analyses for the mass eigenvalue $m$, width $\Gamma$, lifetime
$\tau$, maximal probability $P_{max}$, and mass spectra of the massive KK fermions, we
give an analysis about the effects of a free parameter $V_0$ of the brane on the
fermion resonances. Some new and interesting results are obtained and discussed.

The organization of this paper is as follows. In
Sec.~\ref{secReview}, we review the general aspects of the thick brane model with a piecewise warp factor. The equations of motion and the
solution of the background scalar field are obtained. Besides,
taking into account the interaction between the fermion and the
scalar field by a generalized Yukawa coupling, the
Schr\"{o}dinger-like equations are derived with the KK
decomposition. In Sec.~\ref{resonances}, we provide a complete
analysis of the fermion resonances on the thick brane in
detail. Finally, our discussions and conclusions are presented in Sec.~\ref{secConclusion}.

\section{Review of the thick brane with a piecewise warp factor}
\label{secReview}

In this section, we will review the construction of the thick brane with a piecewise warp factor. With the actions as the starting points, the field equations for the scalar
field and the Dirac field are obtained, respectively. Especially, inspired by the warp factor in \cite{0801.0801}, we construct a new warp factor and the solutions of the scalar field and the
scalar potential are obtained.

\subsection{The setup of the thick brane with a piecewise warp factor}\label{secsetup}

Let us consider the single scalar field thick brane in (4+1)-dimensional space-time
with a scalar potential $V(\phi)$. Specifically, the action describing such a system
is given by
\begin{eqnarray}
 S = \int d^{5}x \sqrt{-g}
  \left[ \frac{1}{2\kappa_{5}^{2}}R -\frac{1}{2}g^{MN}\partial_{M}\phi\partial_{N}\phi
  -V(\phi) \right]\,,~~
 \label{scalarAction}
\end{eqnarray}
where $R$ is the scalar curvature, $g = \det(g_{MN})$, $M,N = 0,1,2,3,4$, and
$\kappa_{5}^{2}=8\pi G_{5}$ with $G_{5}$ the five-dimensional Newtonian constant of
gravitation. We choose the conformally flat metric and the line-element is assumed
as
\begin{eqnarray}
 ds^{2} =g_{MN}dx^{M}dx^{N}
        =e^{2A(z)}(\eta_{\mu\nu}dx^{\mu}dx^{\nu}+dz^{2}), \label{metric}
\end{eqnarray}
where $e^{2A}$ is the warp factor, the metric tensor presents
signature in the form $(-,+,+,+,+)$, and $z$ stands for the
coordinate of the extra space dimension. In the ansatz, the warp
factor and the scalar field are considered to be functions of $z$
only, i.e., $A = A(z)$, $\phi = \phi(z)$. It is straightforward to obtain the Ricci tensor and the curvature scalar:
\begin{eqnarray}
R_{\mu\nu} &=& -(3A^{\prime2}+A^{\prime\prime})\eta_{\mu\nu}\,,\quad R_{44}=-4A^{\prime\prime}\,,   \label{Ricci_tensor} \\
R &=& -4(3A^{\prime2}+2A^{\prime\prime})e^{-2A}\,,  \label{Ricci_scalar}
\end{eqnarray}
where the prime denotes derivative with respect to the fifth coordinate $z$.

The Euler-Lagrange equations can be obtained from the action (\ref{scalarAction})
with the metric ansatz (\ref{metric}). After some simple variational calculations,
the following coupled nonlinear differential equations are obtained:
\begin{eqnarray}
 \phi^{\prime2} &=& 3(A^{\prime2}-A^{\prime\prime})\,,   \label{2ndOrderEqs1} \\
 V(\phi) &=& \frac{3}{2}(-3A^{\prime2}-A^{\prime\prime})e^{-2A}\,,  \label{2ndOrderEqs2} \\
 \frac{d V(\phi)}{d\phi} &=& (3A^{\prime}\phi^{\prime}+\phi^{\prime\prime})e^{-2A} \,. \label{PotentialV1}
\end{eqnarray}

\begin{figure*}
 \subfigure[$e^{2A(z)}$]{{\label{figwarp}}
  \includegraphics[width=0.4\textwidth]{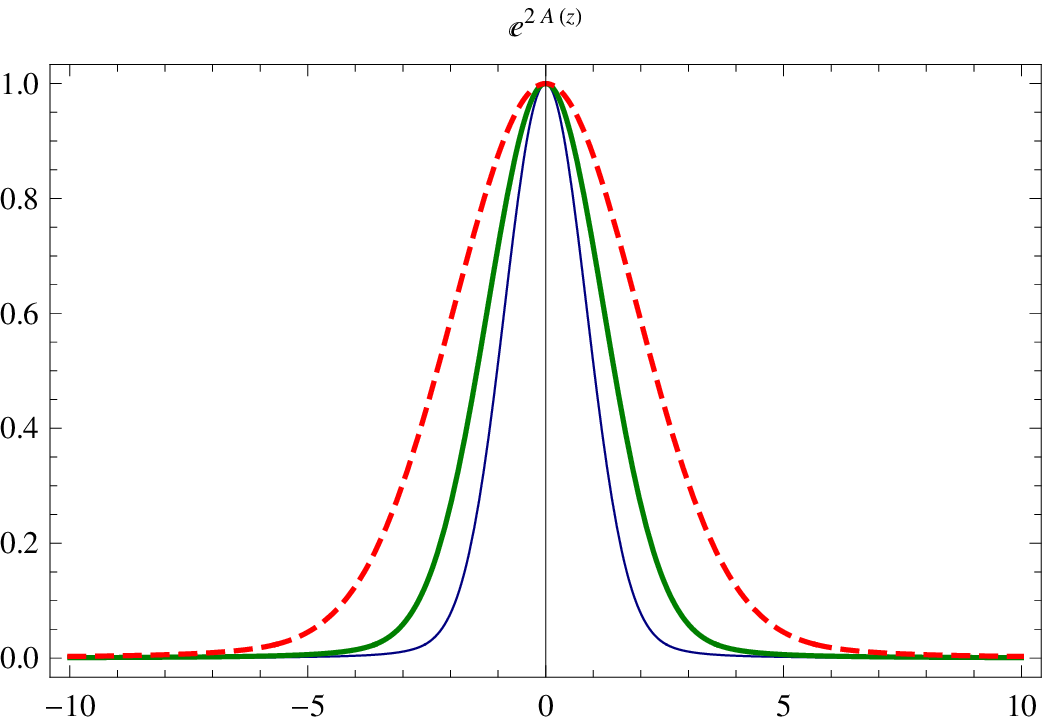}}
    \hspace{1.0cm}
 \subfigure[$R(z)$]{{\label{figricciscalar}}
  \includegraphics[width=0.4\textwidth]{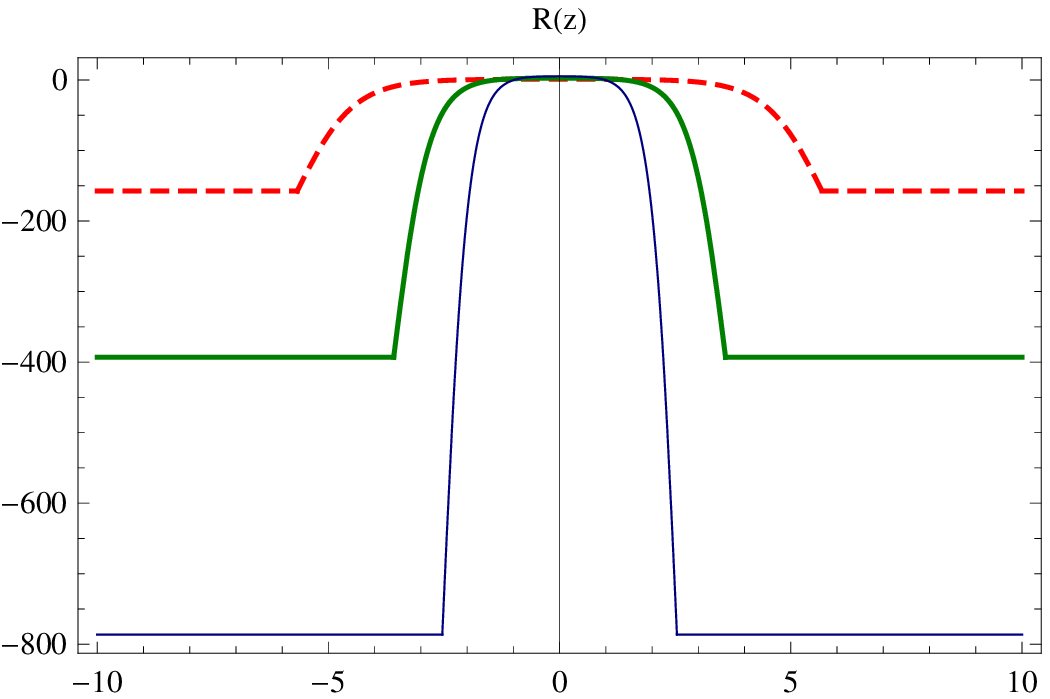}}
\hspace{1.0cm}
 \subfigure[$\phi(z)$]{{\label{figphi}}
  \includegraphics[width=0.4\textwidth]{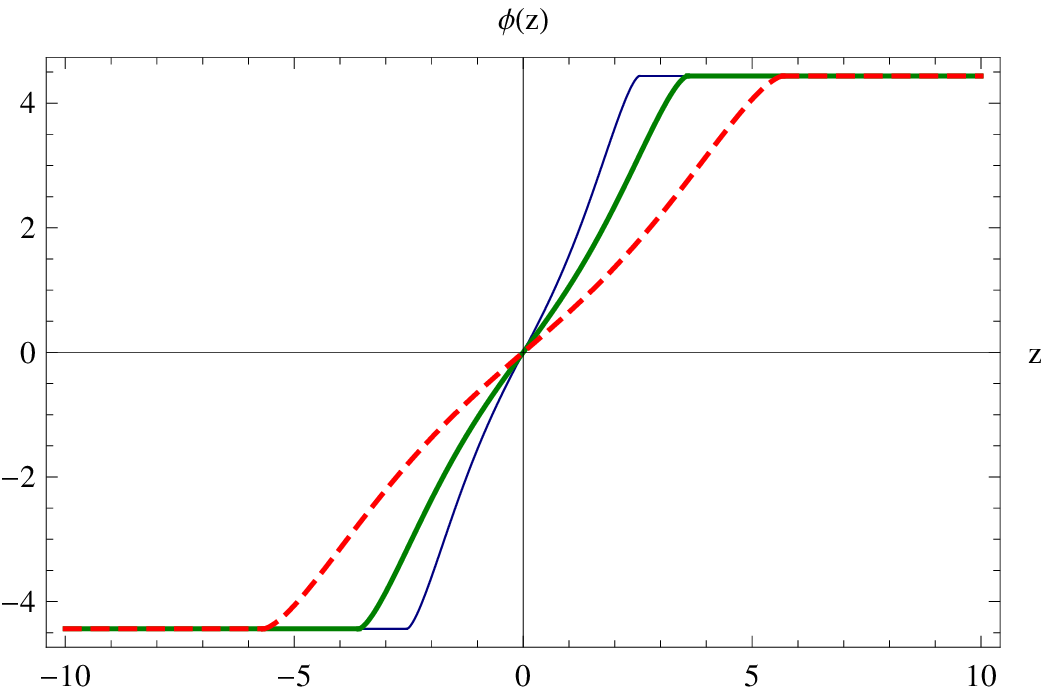}}
\hspace{1.0cm}
 \subfigure[$\phi'(z)$]{{\label{figphiprime}}
  \includegraphics[width=0.4\textwidth]{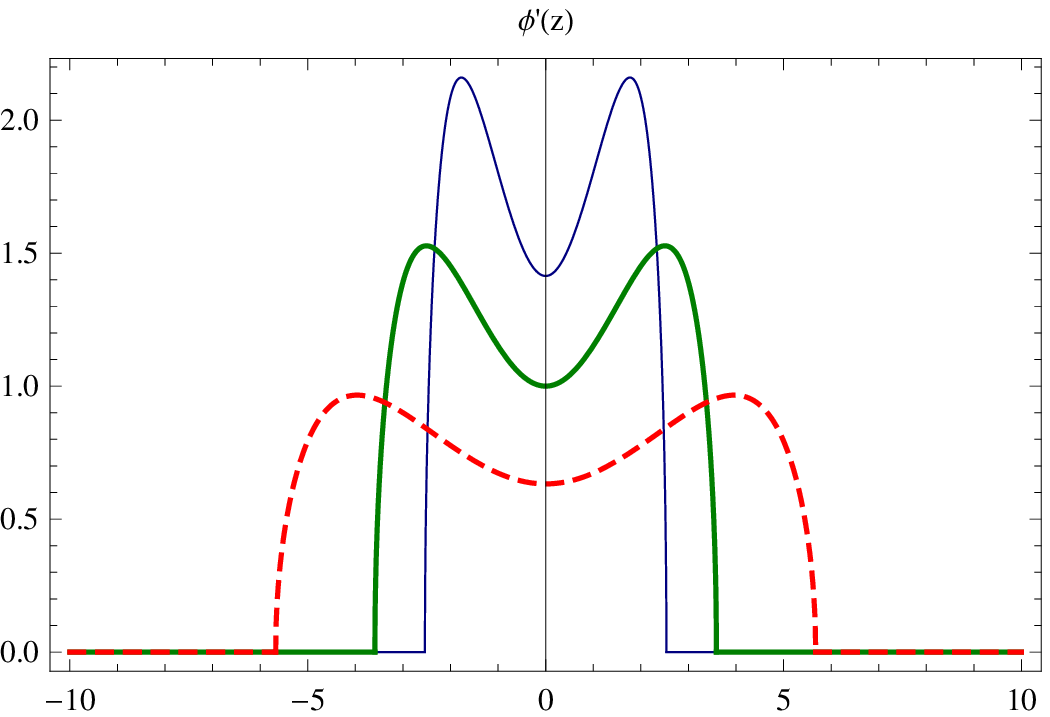}}
\caption{The warp factor $e^{2A(z)}$, curvature scalar $R(z)$, scalar field  $\phi(z)$, and its first derivative $\phi'(z)$. The parameters are
set to $V_0=0.2$ (dashed thicker), $0.5$ (solid thicker), and $1$ (solid thinner).}
 \label{fig_scalar_warpfactor}
\end{figure*}

In order to obtain a thick brane, we modify the warp factor in \cite{0801.0801} as follows:
\begin{eqnarray}
A(z) = \left\{ \begin{array}{ll}
2\ln\big[\frac{\cos(\sqrt{V_{0}}|z|)+2}{3}\big] & \textrm{\,, $|z|\leq\frac{d}{2}$}\\
& \textrm{}\\
-\frac{1}{2}\ln\big[k_{0}^{2}(|z|+\beta)^{2}\big] &
\textrm{\,, $|z|>\frac{d}{2}$}
\end{array} \right. \,,\label{warpA}
\end{eqnarray}
where $V_{0}$, $d$, and $\beta$ are constants, $k_{0}=\sqrt{-\Lambda/6}$
\cite{BehrndtCvetic} and the warp factor is a piecewise defined function. For various values of $V_{0}$, the corresponding warp factor is plotted in Fig.~\ref{figwarp}. We note that the shape of the warp factor here is a normal distribution configuration like in many other thick brane models. It is
worth noting that the continuity of the scalar field $\phi$ and its first order derivative $\phi'$ closely depends on the continuity of the warp factor $A$, its first order derivative $A'$, and its second order derivative $A''$ at the junctions $z=\pm d/2$, which impose the following three junction conditions among the parameters $\beta$, $k_{0}$, $V_0$, and $d$:
\begin{eqnarray}
k_{0}\left(\frac{d}{2} +\beta\right)&=& 9\left[\cos \left(\frac{\sqrt{V_0}
d}{2}\right)+2\right]^{-2}\,, \label{matching1}\\
 \left(\frac{d}{2}+\beta\right)^{-1} &=&\frac{2\sqrt{V_0}\sin
\left(\frac{\sqrt{V_0}d}{2}\right)}{2+\cos \left(\frac{\sqrt{V_0}
d}{2}\right)}\,,\label{matching2}\\
 \left(\frac{d}{2}+\beta\right)^{-2} &=&-\frac{2V_0\left[1+2\cos \left(\frac{\sqrt{V_0}
d}{2}\right)\right]}{\left[2+\cos \left(\frac{\sqrt{V_0}
d}{2}\right)\right]^{2}}\,.\label{matching3}
\end{eqnarray}
From Eqs.~(\ref{matching1}), (\ref{matching2}) and (\ref{matching3}), the parameters $\beta$, $k_{0}$, and $d$ can be derived in terms of $V_0$, namely,
\begin{eqnarray}
d=\frac{2\arccos(\frac{1-\sqrt{7}}{2})}{\sqrt{V_0}},\; \beta=-\frac{1.501~45}{\sqrt{V_0}},\; k_{0}=6.270~8\sqrt{V_0}\,.\nonumber\\
\end{eqnarray}
The above three conditions can ensure the correct Israel junction conditions. The similar situation was explained in the work \cite{0801.0801}.
This means that, in order to get a continuous scalar field without derivative jumps, the
parameters must satisfy the three junction conditions. Thus, with the three junction conditions, we obtain a thick brane, and the thickness of the brane $d$ is characterized by a free parameter $V_0$. In the thick brane scenario, the warp factor is generally expressed in terms of a ``smooth function'' $e^{2A(z)}$ of the fifth coordinate $z$. Here, the second order derivatives of the warp factor are continuous at two junctions $z=\pm d/2$ but the warp factor has third order derivative jumps. In a sense, we can interpret $e^{2A(z)}$ as a smooth curve\cite{smoothness}. So, we call this kind of thick brane the thick brane with a piecewise warp factor matching smoothly. Notice that the junction points $z=\pm d/2$ are located within the first period of the periodic function $\left[({\cos(\sqrt{V_{0}}|z|)+2})/{3}\right]^4$. In this paper, we focus on the problem of the fermion resonances of the massive chiral KK modes in the background of the thick brane with a piecewise warp factor matching smoothly.

From the field equations~(\ref{2ndOrderEqs1}), (\ref{2ndOrderEqs2}), (\ref{PotentialV1}) and the warp
factor~(\ref{warpA}), together with the junction conditions,
the scalar field and the scalar potential are given by:
\begin{eqnarray}
\phi(z) &=& \left\{
\begin{array}{ll}
2i\sqrt{2q}\;\Phi(z)
   & \textrm{, $|z|\leq \frac{d}{2}$}\\
   \text{sign}(z)2i\sqrt{2q}\;\Phi(d/2)
   & \textrm{, $|z|>\frac{d}{2}$}
\end{array}
 \right. ,\label{scalar}\\
\Phi(z)&=&\text{F}\left(\varphi(z),p\right)
-6\,\text{$\Pi$}\left(q/3,\varphi(z),p\right)
\nonumber\\&&+4\,\text{$\Pi$}\left(q,\varphi(z),p\right), \\
\varphi(z)&=&i\text{arcsinh}
  \left[\frac{1}{\sqrt{q}}\tan(\frac{\sqrt{V_{0}}z}{2})\right],\;\nonumber\\
p&=&(-53+20\sqrt{7})/3, \;\;\;\;\; q=2\sqrt{7}-5,\\
 V(\phi(z)) &=& \left\{
 \begin{array}{ll}
       \frac{243V_{0}\left[3\cos(2\sqrt{V_{0}}z)+2\cos(\sqrt{V_{0}}z)-2\right]}{\left[\cos(\sqrt{V_{0}}z)+2\right]^{6}}
     & \textrm{, $|z|\leq \frac{d}{2}$}\\
     & \textrm{}\\
      -6k_{0}^2 & \textrm{, $|z|>\frac{d}{2}$}
\end{array} \right.,\label{PotentialV}\nonumber\\
\end{eqnarray}
where $i$ is the imaginary unit, $\text{F}\left(\varphi(z),p\right)$ is the elliptic integral of the first kind and $\text{$\Pi$}\left(q,\varphi(z),p\right)$ is the incomplete elliptic integral. The shape of the scalar field is shown in Fig.~\ref{figphi}. From
the figure, we can see that the scalar field has a kinklike
structure. Within the intervals $|z|>d/2$,
the scalar field tends to two constants. It is worth noting that we can ensure that the first order derivative $\phi'$ is zero when $|z|\geq d/2$ [see Fig.~\ref{figphiprime}]. In other words, the extradimensional field profile $\phi(z)$ has no derivative jumps at $z=\pm d/2$. It is easy to check that the curvature scalar $R(z)$ and the scalar potential $V(\phi(z))$ are continuous~\footnote{We thank the referee for pointing out the problem of the discontinuity of the scalar potential in the first version of the paper.}.

Here, the energy density of the thick brane is
\begin{eqnarray}
 T_{00} &=& g_{00}\mathcal {L}-2\frac{\partial\mathcal {L}}{\partial
g^{00}}\nonumber \\
 &=& e^{2A(z)}\left(\frac{1}{2}\phi^{\prime2}+V(\phi)\right)\,, \label{Energydensity}
\end{eqnarray}
where $\mathcal {L}$ is the Lagrangian density of the matter field. The distribution of the energy density $T_{00}$ along the extra dimension $z$ is
illustrated in Fig.~\ref{Energydensity}. The energy density has no double-peaked structure which is seen as characterizing a brane with internal structure~\cite{Bazeia:03a,Almeida:0901}.

\begin{figure}
  \includegraphics[width=0.4\textwidth]{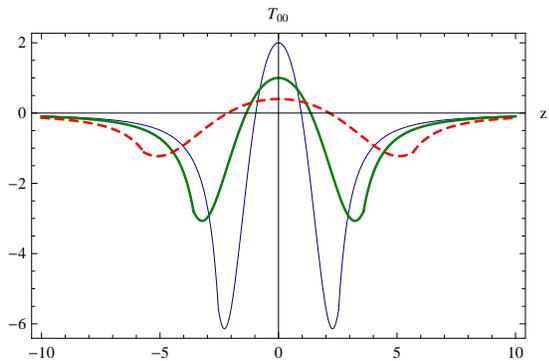}
\caption{The energy density $T_{00}(z)$. The parameters are
set to $V_0=0.2$ (dashed thicker), $0.5$ (solid thicker), and $1$ (solid thinner).}
\label{Energydensity}
\end{figure}

We now want to argue that, the massless mode of gravity (i.e., the four-dimensional graviton) can be localized on the brane. We consider the gravitational perturbation with the metric fluctuations given by
$ds^{2}=e^{2A(z)}((\eta_{\mu\nu}+h_{\mu\nu}(x,z))dx^{\mu}dx^{\nu}+dz^{2})$. Furthermore, we can decompose $h_{\mu\nu}(x,z)$ in the form $h_{\mu\nu}=e^{\frac{3}{2}A}\hat{h}_{\mu\nu}(x)\hat{\psi}_m(z)$, and require that $\hat{h}$ be a four-dimensional mass eigenstate mode $\Box \hat{h}_{\mu\nu}=m^{2}\hat{h}_{\mu\nu}$, where $\Box=\eta^{\mu\nu}\partial_\mu\partial_\nu$. The Schr\"{o}dinger-like equation under the transverse-traceless gauge can be obtained\cite{Randall:99b,DeWolfe:00a,Csaki:00a,BehrndtCvetic,SkenderisChamblin}:
\begin{eqnarray}
-\hat{\psi}_{m}''(z)+V_\text{G}(z)\hat{\psi}_{m}(z)=m^{2}\hat{\psi}_{m}(z)\,,\label{grascheq}
\end{eqnarray}
with the effective potential
\begin{eqnarray}
V_\text{G}(z)=\frac{3}{2}A''+\frac{9}{4}A'^{2}\,.\label{grapotential}
\end{eqnarray}

With the warp factor~(\ref{warpA}), it is easy to check that this is a volcano potential like in the RS model. Furthermore, there exists a zero mode solution $\hat{\psi}_0\propto e^{\frac{3}{2}A(z)}$. It can be shown that the graviton zero mode actually is normalizable, namely, $\int_{-\infty}^{+\infty}dz |\hat{\psi}_0|^{2}\propto c_0+k_0^{-3}(d/2+\beta)^{-2}$ with $c_0=\frac{2}{729}\int_{0}^{d/2}dz[\cos(\sqrt{V_{0}}|z|)+2]^{6}$ a constant. So the massless four-dimensional graviton can be produced by a zero mode\cite{Randall:99b,DeWolfe:00a,Gremm:00a,Csaki:00a,WangPRD2002}.

Here, the observed four-dimensional effective Planck scale $M_{Pl}$ can be derived. As we know, the five-dimensional Einstein-Hilbert action of gravity can be reduced to the four-dimensional one, namely,
\begin{eqnarray}
S&\sim&M_\ast^3\int dx^5\sqrt{g^{(5)}}R^{(5)}\nonumber\\
&\supset& M_\ast^3\int e^{3A} dz\int dx^4\sqrt{g^{(4)}}R^{(4)}\nonumber\\
&=& M_{Pl}^2\int dx^4\sqrt{g^{(4)}}R^{(4)}\,,~~~\label{5Dgraaction}
\end{eqnarray}
where $M_\ast$ is the five-dimensional fundamental scale. Thus the four-dimensional effective Planck scale can be derived:
\begin{eqnarray}
M_{Pl}^2&=&M_\ast^3\int^{+\infty}_{-\infty}e^{3A} dz\nonumber\\
&=&M_\ast^3[c_0+k_0^{-3}(d/2+\beta)^{-2}]\,.~~\label{4Dgraaction}
\end{eqnarray}
The four-dimensional effective Planck scale can be expressed through the fundamental scale and a finite one-dimensional warped volume.

\subsection{Spin 1/2 fermion fields on the brane}\label{secSpin1-2}

In this subsection, we will review the general aspects of the spin half fermions on
the brane. Let us consider a massless bulk fermion coupled to the background scalar
by the generalized Yukawa coupling in five-dimensional space-time. The action reads
\begin{eqnarray}
 S_{1/2} &=& \int d^{5}x\sqrt{-g}\left[\bar{\Psi}\Gamma^{M}D_{M}\Psi
    -\eta\bar{\Psi}F(\phi)\Psi\right],
    \label{fermion field action}
\end{eqnarray}
where the spin connection $\omega_{M}$ in the covariant derivative
\begin{eqnarray}
D_{M}\Psi &=& (\partial_{M}+\omega_{M})\Psi
\label{covariant derivative}
\end{eqnarray}
is defined as
\begin{eqnarray}
\omega_{M}=\frac{1}{4}\omega^{\bar{M}\bar{N}}_{M}
         \Gamma_{\bar{M}}\Gamma_{\bar{N}}\,.
   \label{SpinConnection}
\end{eqnarray}
In five-dimensional space-time, the spinor structure of the four component fermions
is determined by $\Gamma^{M}=E^{M}_{\bar{M}}\Gamma^{\bar{M}}$ with the
$E^{M}_{\bar{M}}$ being the vielbein and $\{\Gamma^{M},\Gamma^{N}\}=2g^{MN}$. Here
we use the capital Latin letters $M,N,\ldots$ and $\bar{M},\bar{N},\ldots$ to label
the indices of five-dimensional space-time coordinates and the local Lorentz ones,
respectively. And $\Gamma^{M}$ are the curved gamma matrices and $\gamma^{\bar{M}}$
are the flat ones.

With the conformally flat metric (\ref{metric}), the nonvanishing
components of the spin connection $\omega_{M}$ are given by
\begin{eqnarray}
\omega_{\mu}=\frac{1}{2}(\partial_{z}A)\gamma_{\mu}\gamma_{5}\,.
\label{SpinConnectionOmegaMu}
\end{eqnarray}
Then the five-dimensional Dirac equation can be derived from the action
(\ref{fermion field action})
\begin{eqnarray}
\left[\gamma^{\mu}\partial_{\mu}+\gamma^{5}\left(\partial_{z}+2\partial_{z}A\right)-\eta
e^{A}F(\phi)\right]\Psi=0\,,
 \label{DiracEq1}
\end{eqnarray}
where $\gamma^{\mu}\partial_{\mu}$ is the four-dimensional Dirac
operator on the brane.

In what follows, we focus on the fermion field equations of the
five-dimensional fluctuations, and reduce the Dirac fermion field
$\Psi$ to the four-dimensional effective field on the brane.
Following the routine in Ref.~\cite{Liu:09a}, we carry out the general KK
decomposition according to the chiralities of the fermions,
\begin{eqnarray}
\Psi(x,z)=e^{-2A}\sum_{n}\left[\psi_{\text{L}n}(x)f_{\text{L}n}(z)
+\psi_{\text{R}n}(x)f_{\text{R}n}(z)\right]\,,\nonumber\\
\label{the general chiral decomposition}
\end{eqnarray}
where $\psi_{\text{L}n}(x)=-\gamma^{5}\psi_{\text{L}n}(x)$ and
$\psi_{\text{R}n}(x)=\gamma^{5}\psi_{\text{R}n}(x)$ are the left-handed and right-handed
components of the Dirac fermion fields on the brane, respectively. They satisfy the
four-dimensional massive Dirac equations in the form of
$\gamma^{\mu}\partial_{\mu}\psi_{\text{L}n}(x)=m_{n}\psi_{\text{R}n}(x)$ and
$\gamma^{\mu}\partial_{\mu}\psi_{\text{R}n}(x)=m_{n}\psi_{\text{L}n}(x)$. At the same time, the KK
modes $f_{\text{L}n}(z)$ and $f_{\text{R}n}(z)$ of the chiral decomposition about the Dirac fields
meet the form of the following coupled equations:
\begin{eqnarray}
 \left[\partial_{z} +{\eta} e^{A}F(\phi)\right]
  f_{\text{L}n}(z) &=& \,\,\,\,\,m_{n}f_{\text{R}n}(z)\,,\label{spinorCoupledEqs1} \\
 \left[\partial_{z}-\eta e^{A}F(\phi)\right]
  f_{\text{R}n}(z) &=& -m_{n}f_{\text{L}n}(z)\,. \label{spinorCoupledEqs2}
\end{eqnarray}
With the purpose of obtaining the standard four-dimensional
effective action for the massive chiral fermions, the following
orthonormality conditions for $f_{\text{L}n}(z)$ and $f_{\text{R}n}(z)$ are
needed:
\begin{eqnarray}
 \int^{+\infty}_{-\infty}f_{\text{L}m}f_{\text{L}n}dz&=&
 \int^{+\infty}_{-\infty}f_{\text{R}m}f_{\text{R}n}dz=\delta_{mn}\,,\nonumber\\
 \int^{+\infty}_{-\infty}f_{\text{L}m}f_{\text{R}n}dz&=&0\,.
  \label{OrthonormalityConditions}
\end{eqnarray}
Inspecting of Eqs.~(\ref{spinorCoupledEqs1}) and
(\ref{spinorCoupledEqs2}), we have
\begin{eqnarray}
 \left(\frac{d}{dz}-\eta e^{A}F(\phi)\right)
       \left(\frac{d}{dz}+\eta e^{A}F(\phi)\right)f_{\text{L}}
    =-m^{2}_{n}f_{\text{L}}\label{ScheqLeftLeft}\,,~~~\\
 \left(\frac{d}{dz}+\eta e^{A}F(\phi)\right)
       \left(\frac{d}{dz}-\eta e^{A}F(\phi)\right)f_{\text{R}}
    =-m^{2}_{n}f_{\text{R}}\label{ScheqRightRight} \,.~~~
\end{eqnarray}
We have thus obtained the Schr\"{o}dinger-like equations for the left
and right chiral fermions \cite{Almeida:0901,Liu:09b}
\begin{eqnarray}
H_{\text{L}}f_{\text{L}}(z)=m^{2}f_{\text{L}}(z)\,,\\
H_{\text{R}}f_{\text{R}}(z)=m^{2}f_{\text{R}}(z)\,,
\end{eqnarray}
where the corresponding Hamiltonians are defined as
\begin{eqnarray}
 H_{\text{L}}=\left(-\frac{d}{dz}+\eta e^{A}F(\phi)\right)
  \left(\frac{d}{dz}+\eta e^{A}F(\phi)\right)\,, \label{HamiltonianLeft}\\
 H_{\text{R}}=\left(-\frac{d}{dz}-\eta e^{A}F(\phi)\right)
 \left(\frac{d}{dz}-\eta e^{A}F(\phi)\right)\,.
          \label{HamiltonianRight}
\end{eqnarray}
In view of (\ref{ScheqLeftLeft}) and (\ref{ScheqRightRight}), it may
seem obvious that the Schr\"{o}dinger equations can be rewritten as
\begin{subequations}\label{Scheq}
\begin{eqnarray}
 \left[-\partial_{z}^{2}+V_{\text{L}}(z)\right]f_{\text{L}} &=& m^{2}f_{\text{L}} \,,
    \\ \label{ScheqLeft}
 \left[-\partial_{z}^{2}+V_{\text{R}}(z)\right]f_{\text{R}} &=& m^{2}f_{\text{R}} \,,
       \label{ScheqRight}
\end{eqnarray}
\end{subequations}
where the effective potentials of the KK modes have the explicit expressions
\begin{subequations}\label{Vz}
\begin{eqnarray}
V_{\text{L}}(z)&=&[\eta e^{A}F(\phi)]^{2}-\partial_{z}[\eta
e^{A}F(\phi)]\,, \label{VzL}  \\
V_{\text{R}}(z)&=&[\eta e^{A}F(\phi)]^{2}+\partial_{z}[\eta
e^{A}F(\phi)]\,.
\label{VzR}
\end{eqnarray}
\end{subequations}

\section{Fermion resonances on the thick brane}
\label{resonances}

In this section, we mainly discuss the problem of the fermion resonances on the thick brane with a piecewise warp factor. Specifically, we investigate the effects of two
parameters $k$ and $V_0$ on the resonant states of the massive KK modes, respectively. Here
$k$ is a parameter in the generalized Yukawa coupling
$\eta\overline{\Psi}\phi^{k}\Psi$, $k=1,3,5,\ldots$. Our discussions focus on the
properties, i.e., the mass spectra, the lifetimes, and the number of the massive resonant
KK modes, related to the two parameters.

In Ref.~\cite{0801.0801}, the parameter $x$ which effectively parametrizes the thickness of the domain wall has a very interesting impact on
the gravitation resonances. In this paper, we want to know the impacts of the
parameters on the behaviors of the fermion resonances.

In this section, some new natures of the resonant states are obtained. We will show that the number of the resonant states increases when $k$ becomes larger, however, it decreases as $V_0$ increases. The height of  the potential function of the resonant fermions with left-handed chirality is the same as that of the one with right-handed chirality. The parameter $V_0$ determines the depth of the potential well of the resonant fermions, but it has no influence on the height of the potential function.

\subsection{The fermion zero modes}\label{zeromodes}

As we know, in order to study the localization and resonance (quasilocalization ) problems, the coupling between the fermion
field and the background scalar field should be introduced. Here, we choose the generalized Yukawa
coupling $F(\phi)=\phi^{k}$ with odd $k=$ $1$, $3$, $5$, $\ldots$.
In this subsection, we mainly study the localization of the zero modes of left- and right-handed fermions on the brane.

With the given $k=1$ and brane solution $\phi(z)$ in Eq.~(\ref{scalar}) obtained in
the previous section, the potential functions can be rewritten in the following
specific forms:
\begin{widetext}
\begin{subequations}\label{VzCaseI}
\begin{eqnarray}
V_{\text{L}}(z) &=& \left\{ \begin{array}{ll}
\frac{\eta l}{81} \left[ \eta l^{3}\phi^{2}(z)+18\phi(z)\sqrt{V_0}\sin(\sqrt{V_0z})
+\frac{9q\sqrt{3V_0}\left[\sec^2\left(\frac{\sqrt{V_0z}}{2}\right)+4l-16\right]}{\sqrt{\left[q+\tan^2\left(\frac{\sqrt{V_0z}}{2}\right)\right]\left(19-7\sqrt{7}+\frac{3m}{l-1}\right)}}\right]& \textrm{\,, $|z|\leq\frac{d}{2}$}\\
& \textrm{}~~~~~~~~\,,\\
\frac{\eta \phi(z) \left(\eta \phi(z) (z\pm\beta )+\sqrt{k_0^2 (z\pm\beta
   )^2}\right)}{k_0^2 (z\pm\beta )^3} & \textrm{\,, $|z|>\frac{d}{2}$}
\end{array} \right.  \\ \label{VzCaseIL}
V_{\text{R}}(z) &=& V_{\text{L}}(z)|_{\eta\rightarrow-\eta}\,,\;\;\;\; \label{VzCaseIR}
\end{eqnarray}
\end{subequations}
\end{widetext}
where $l=2+ \cos(\sqrt{V_0z})$.

\begin{figure*}
 \subfigure[$V_{\text{L}}(z)$]{{\label{figpotentialL}}
  \includegraphics[width=0.4\textwidth]{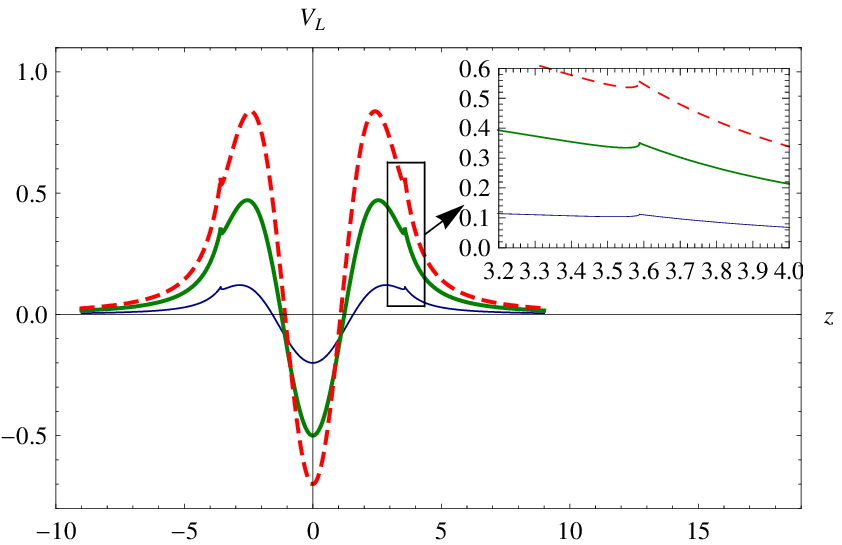} }
 \subfigure[$V_{\text{R}}(z)$]{{\label{figpotentialR}}
  \includegraphics[width=0.4\textwidth]{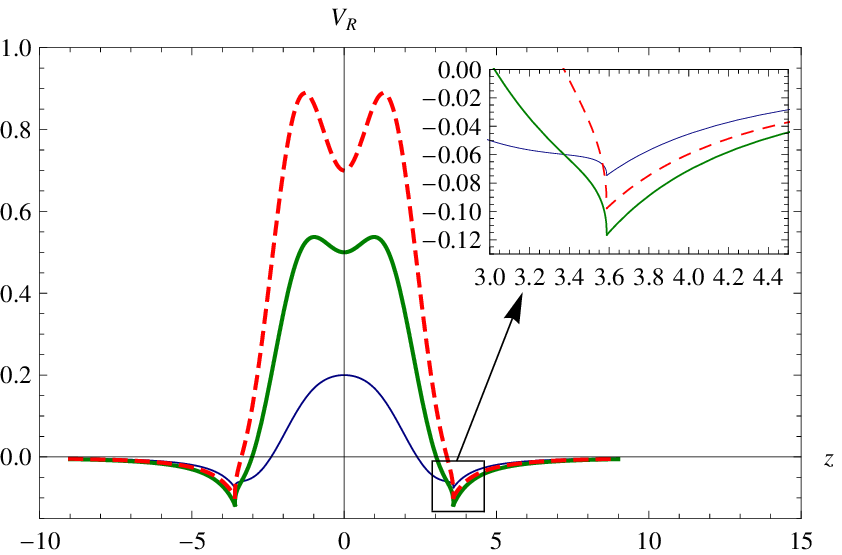}}
\caption{The potential functions $V_{\text{L}}(z)$ and $V_{\text{R}}(z)$ for
$F(\phi)=\phi$. The insets show the enlarged view of the junction points for clarity. The parameters are
set to $V_0=0.5$, $\eta=0.7$ (dashed thicker), $0.5$ (solid thicker), and $0.2$ (solid thinner).}
 \label{figpotential}
\end{figure*}

In what follows, we analyze the asymptotic behavior of the potential functions for
the left- and right-handed fermions. After simple calculation and analysis, we find
that both the potential functions have a very good asymptotic behavior. Specifically,
when $z$ tends to zero and infinity, their asymptotic values are constants. It can
be expressed as follows:
\begin{eqnarray}
V_{\text{L}}(0) = -V_{\text{R}}(0)&=&-\sqrt{2V_{0}}\eta \,,
    \\ \label{asymptoticvaluea}
V_{\text{L},\text{R}}(z\rightarrow\pm\infty)&\rightarrow& 0\,.
       \label{asymptoticvalueb}
\end{eqnarray}
It can be seen that, for the same set of parameters, the potential
function of left-handed fermion KK modes $V_\text{L}(z)$ has an opposite value compared to the one of right-handed fermion KK modes $V_\text{R}(z)$ at the coordinate
origin $z=0$. When $z$ tends to infinity, both potential
functions vanish. The behaviors of the two potential functions
are illustrated in Fig.~\ref{figpotential} for given $V_0$ and various values of $\eta$.
From Eq.~(\ref{Vz}) and the continuity of the first derivatives of the scalar field and warp factor, we know that the potentials $V_{\text{L}}(z)$ and $V_{\text{R}}(z)$ are continuous. However, they are not smooth because the first order derivative of the scalar field is not smooth, this can be seen from Fig.~\ref{figphiprime}. The insets in Fig.~\ref{figpotential} show the enlarged view of the continuity of the potentials for clarity.
From the left panel of Fig.~\ref{figpotential}, we
can see that $V_{\text{L}}(z)$ is a modified volcano-type potential like in most of the other
contexts. So there is no discrete mass spectrum of bound states, and there exists no
mass gap to separate the massless mode from the massive KK modes for $V_{\text{L}}(z)$.

In Fig.~\ref{figpotential}, from the right panel, we see that there is a potential barrier around the
brane location for $V_{\text{R}}(z)$. Nevertheless, for stronger coupling (bigger $\eta$),
the potential well emerges at the top of the barrier. This means that the behavior of the potential function is related to the scalar-fermion coupling. In order to clearly examine
the impact of $V_0$ on the behavior of $V_{\text{R}}(z)$, we fix
$k$ and $\eta$ in the following. The shapes of $V_{\text{L}}(z)$ and $V_{\text{R}}(z)$ for various values of $V_0$ can be  depicted easily. As we know, for $V_{\text{R}}(z)$, in general, the potential barrier can not trap a massive fermion. However, when $V_0$ becomes smaller, the potential well located at the top of $V_{\text{R}}(z)$ is deep enough, and a massive fermion with a finite lifetime would appear, however, the height of the potential remains unchanged. This means that the phenomenon of massive resonant states will also occur.

If we recast the scalar-fermion coupling $\eta\bar{\Psi}\phi\Psi$ ($k=1$) to
$\eta\bar{\Psi}\phi^{3}\Psi$ ($k=3$), the configurations of the two potential functions $V_{\text{L}}(z)$ and $V_{\text{R}}(z)$ will change. In this situation, $V_{\text{L}}(z)$ becomes a double-well potential,
but the right-handed one is still a single-well potential. However, for the two potential functions, if the other parameters are fixed, when the
parameter $V_0$ becomes smaller, the potential well will become deeper. Furthermore, at
the center of the brane along the direction of the extra dimension, the values of
the potential functions $V_{\text{L},\text{R}}(z)$ are zero, i.e., $V_{\text{L}}(0)$ $=$ $V_{\text{R}}(0)$ $=$
$0$. This result is similar to that obtained in Refs.~\cite{Liu:09b,Liu:09d}.

\begin{figure}
\includegraphics[width=0.4\textwidth]{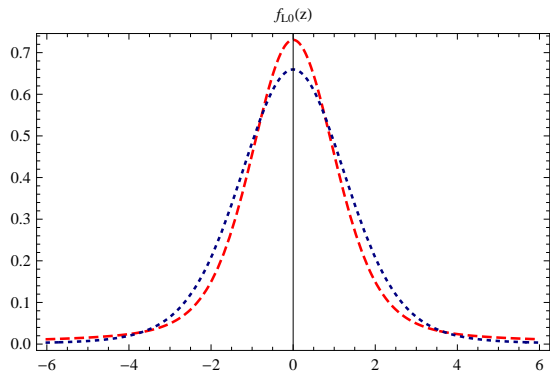}
\caption{The normalizable fermion zero mode $f_{\text{L0}}(z)$. The parameters are set to $\eta=1.0$, $V_0=0.2$ for the dotted line and $V_0=0.5$ for the dashed line.} \label{zeromode}
\end{figure}

As discussed above, $V_{\text{L}}(z)$ is always a modified volcano type of potential.
It is well known that the potential of the left-handed fermions does not provide
mass gap between the zero mode and KK excitation modes, so there is a continuous
gapless spectrum of KK excitation modes. From Eqs.~(\ref{spinorCoupledEqs1}) and
~(\ref{spinorCoupledEqs2}), the zero mode of the left-handed fermions is derived as
\begin{eqnarray}
 &&f_{\text{L}0}(z)  \propto  \exp\left(-\eta\int_{0}^{z}d\bar{z}e^{A(\bar{z})}
       \phi(\bar{z})\right)\nonumber\\
&=& \left\{ \begin{array}{ll}
\exp\left(-\frac{\eta}{9}\int_{0}^{z}d\bar{z}[\cos(\sqrt{V_{0}}|\bar{z}|)+2]^{2}\phi(\bar{z})\right), & \textrm{ $|z|\leq\frac{d}{2}$}\\
& \textrm{}\\
\exp\left(-\frac{\eta}{9}\int_{0}^{d/2}d\bar{z}[\cos(\sqrt{V_{0}}|\bar{z}|)+2]^{2}\phi(\bar{z})\right)\nonumber\\
\cdot\exp\left(-\frac{\eta}{k_{0}}\int_{d/2}^{z}d\bar{z}\frac{\phi(\bar{z})}{|\bar{z}|+\beta}\right), &
\textrm{ $|z|\geq\frac{d}{2}$}
\end{array} \right..
       \label{fL0CaseIa}
\end{eqnarray}
When $|z|\geq d/2$, in the first factor, because the integrand and the integrating range are finite, the integration result must be a constant:
\begin{eqnarray}
c_1=\exp\left(-\frac{\eta}{9}\int_{0}^{d/2}d\bar{z}[\cos(\sqrt{V_{0}}|\bar{z}|)+2]^{2}\phi(\bar{z})\right)\,.~~
\end{eqnarray}
In the second factor, we know that $\phi(\bar{z})=c_2$ (a constant) when $\bar{z}\geq d/2$. So, we obtain
\begin{eqnarray}
 f_{\text{L}0}(z) & \propto & \exp\left(-\frac{\eta c_2}{k_{0}}\int_{d/2}^{z}d\bar{z}\frac{1}{|\bar{z}|+\beta}\right)\,,\nonumber\\
 &=& c_3(|z|+\beta)^{-\frac{\eta c_2}{k_{0}}}\,,
       \label{fL0CaseIa2}
\end{eqnarray}
where $c_3=\left(d/2+\beta\right)^{\eta c_2/k_{0}}$. Then the normalization condition
\begin{eqnarray}
\int_{-\infty}^{+\infty}f^2_{\text{L}0}(z) dz =1
\end{eqnarray}
is equivalent to
\begin{eqnarray}
\int_{d/2}^{+\infty}dz(z+\beta)^{-\frac{2\eta c_2}{k_{0}}} < \infty.
      \label{fL0CaseIb}
\end{eqnarray}
It is clear that the normalization condition is turned out to be $2\eta c_2/k_{0}>1$. Hence, the zero mode
(\ref{fL0CaseIa}) is normalizable only if $\eta c_2/k_{0}>1/2$. Besides, although there is no explicit expression for $f_{\text{L}0}(z)$, we can numerically integrate (\ref{fL0CaseIa}). The plot of the normalizable zero mode $f_{\text{L}0}(z)$ via numerical integration is given in Fig.~\ref{zeromode}. From Fig.~\ref{zeromode}, we can better understand the process of the calculation. On the other hand, the potential $V_{\text{R}}(z)$ around the brane location is always positive, and it gradually becomes zero when $z\rightarrow\infty$. We know that this type of potential cannot trap any bound
state fermions with right chirality and there exists no zero mode of right-handed
fermions. This result is consistent with the previous well-known conclusion that massless fermions must be single chirality in the brane world models \cite{Ringeval:02,Liu:09b}.

Nevertheless, for a given $V_0$, the structure of the potential $V_\text{R}$ is determined by
the parameters $k$ and the coupling constant
$\eta$ jointly. For a given $k$, when $\eta$
increases, a potential well around the location of the brane would emerge and the well would be deeper and deeper. In this case, there may exist massive resonant KK modes. In order to
make a potential well appear, we need very large $\eta$, i.e., the coupling between the
scalar and the fermion is very strong~\cite{Liu:09b,Liu:09d}. In
this paper, we would not consider this case and our main task is focusing on the impacts of parameters $k$ and $V_0$ on the massive resonant states.

\subsection{The massive KK modes}\label{seck}

As mentioned above, in this subsection, we mainly discuss the problem of the fermion resonances on the thick brane. Some interesting properties and results will be obtained. Following the routine in Refs.~\cite{Liu:09b,Liu:09d}, with the potential functions~(\ref{VzCaseI}), we get the numerical solutions of the Schr\"{o}dinger
equations~(\ref{Scheq}) with the Numerov algorithm. Furthermore, the resonant probabilities, lifetimes  and mass spectra of the resonances will be obtained.

When the parameter $k$ in scalar-fermion coupling is treated as a variable, the maxima of potential functions $V_{\text{L}}(z)$ and $V_{\text{R}}(z)$ are the same magnitude and the properties of resonances are same to the results obtained in Refs.~\cite{Liu:09b,Liu:09d}. When we treat the parameter $V_0$ as a variable, the maxima of potential functions $V_{\text{L}}(z)$ and $V_{\text{R}}(z)$ are also the same magnitude but the resonant behaviors of the massive chiral fermions with the left- and right-handed chiralities are different from the case in that $k$ is treated as a variable. But the resonant states satisfy the Kaluza-Klein parity-chirality decompositions of massive fermion resonances obtained in Ref.~\cite{Liu:09d}, because the resonant peaks have a one-to-one correspondence between left- and right-handed fermions at some resonant mass eigenvalue.

\subsubsection{Case I: $k$ as a variable}\label{seck}

In this subsection, we mainly study the impacts of the parameter $k$,
which is related to the coupling type between the scalar field and the fermion, on
the massive resonant KK modes.
From the context of numerical analysis, we know that one should
impose the initial or boundary conditions for the second order
differential equations~(\ref{Scheq}). Here, we attach two types of
initial conditions for Eqs.~(\ref{Scheq}). They are
\begin{eqnarray}
f(0)=d_{0},\,\, f'(0)=0 \,, \label{evencondition}
\end{eqnarray}
and
\begin{eqnarray}
f(0)=0,\,\, f'(0)=d_{1} \,. \label{oddcondition}
\end{eqnarray}
Because the potential functions have a $Z_{2}$ symmetry, the solutions of the Schr\"{o}dinger
equations~(\ref{Scheq}) with even
parity and odd parity will be
obtained for the above two conditions, respectively. In this paper, the constants $d_{0}$ and
$d_{1}$ are set to $d_{0}=1$, $d_{1}=5$ for the sake of convenience.
From the point of view of quantum mechanics, we know that
the massive KK modes will feel a combined effect of the potential
barrier and the potential well around the location of the brane. Therefore, the massive KK
modes will show different natures when the depth of the potential
well changes.

\begin{figure*}
 \subfigure[$m^{2}=0.072$ (res.), $0.4378$ (off res.)]{{\label{figunifya}
  \includegraphics[width=0.45\textwidth]{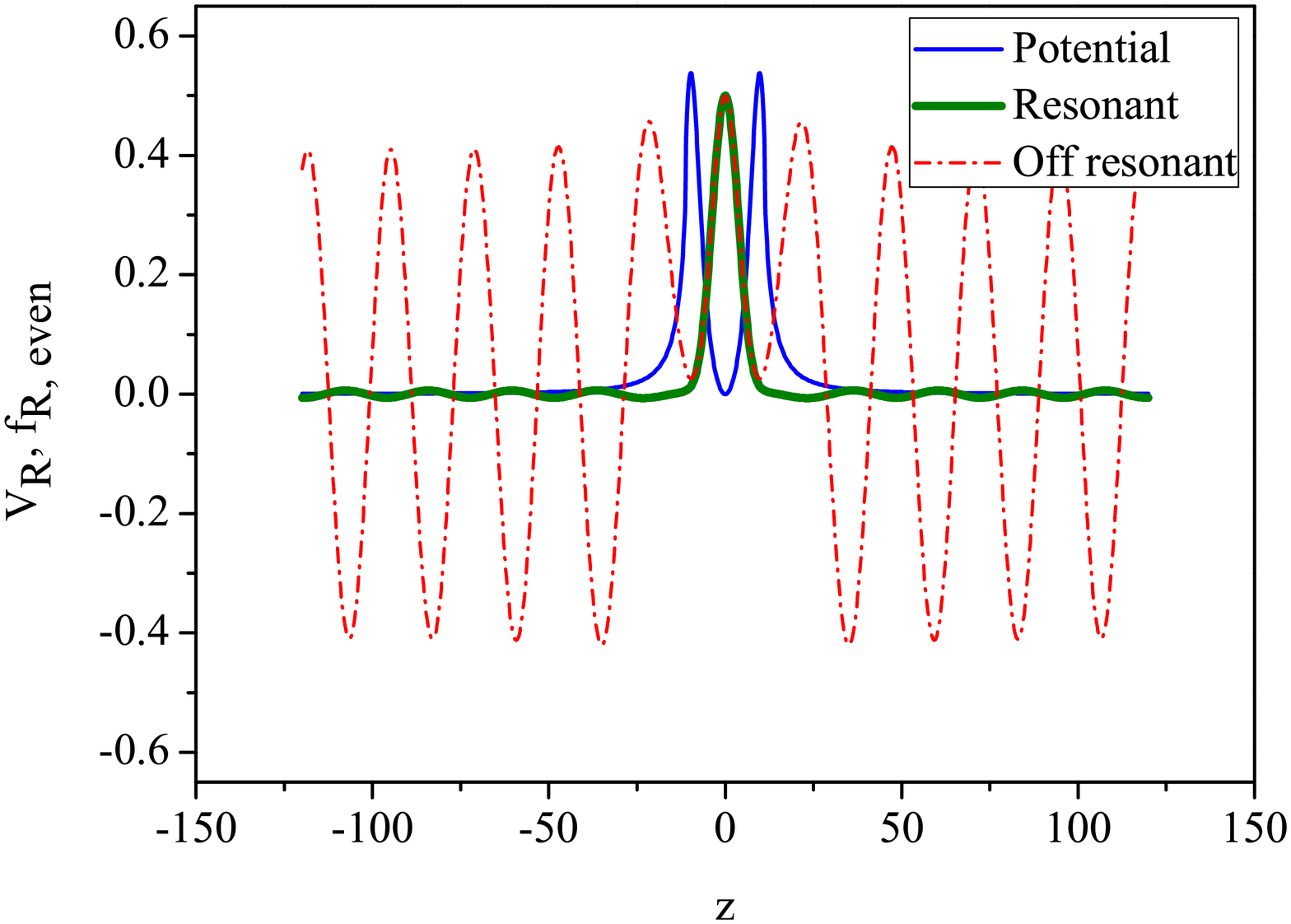}}}
\hspace{1.0cm}
 \subfigure[$m^{2}=0.072$ (res.), $0.439$ (off res.)]{{\label{figunifyb}
  \includegraphics[width=0.45\textwidth]{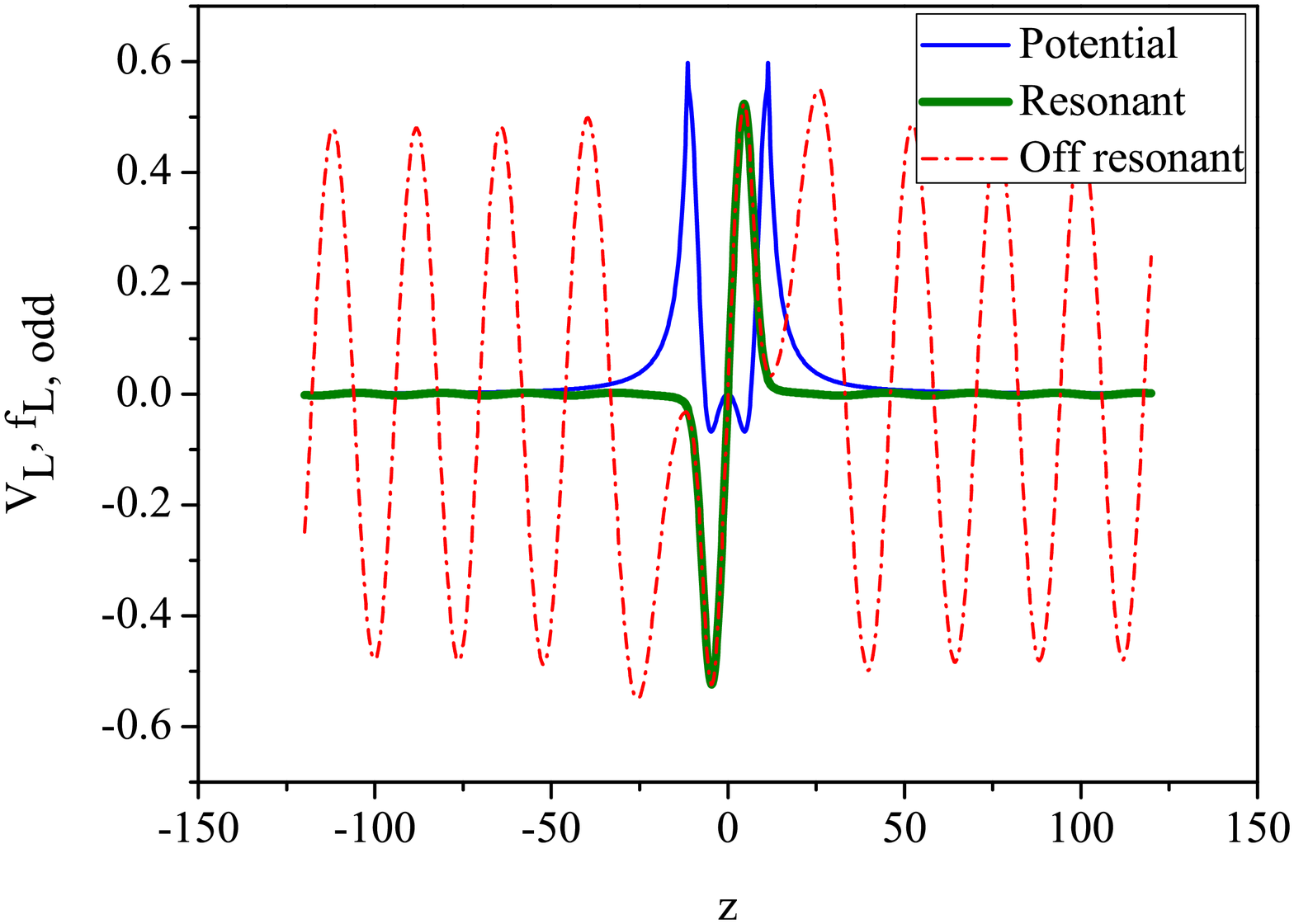}}}
\caption{(color online). The shapes of massive resonant KK modes, potential functions,
off-resonant wave functions of left-handed and
right-handed fermions with even parity and odd parity
for $F(\phi)=\phi^{3}$. The parameters are set to $z_{\text{max}}=120$, $V_0=0.05$, and $\eta=0.05$. The terms ``res." and ``off res." stand for resonant states and off-resonant states, respectively.}
\label{figunifyk=13}
\end{figure*}

\begin{figure*}
 \subfigure[$P_{\text{L}}(m^{2})$]{\label{figPunitya}
  \includegraphics[width=0.45\textwidth]{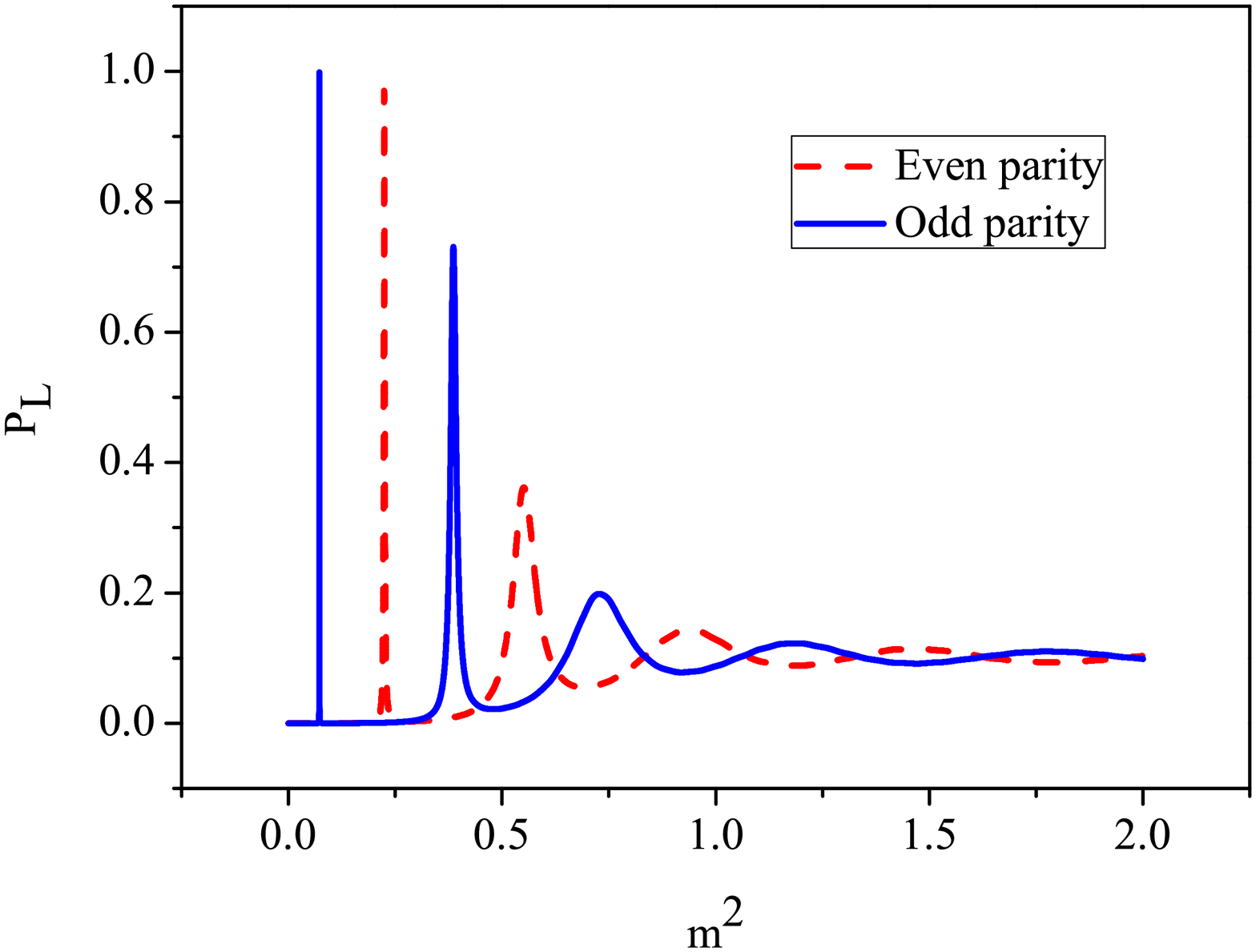}}
\hspace{1.0cm}
 \subfigure[$P_{\text{R}}(m^{2})$] {\label{figPunityb}
  \includegraphics[width=0.45\textwidth]{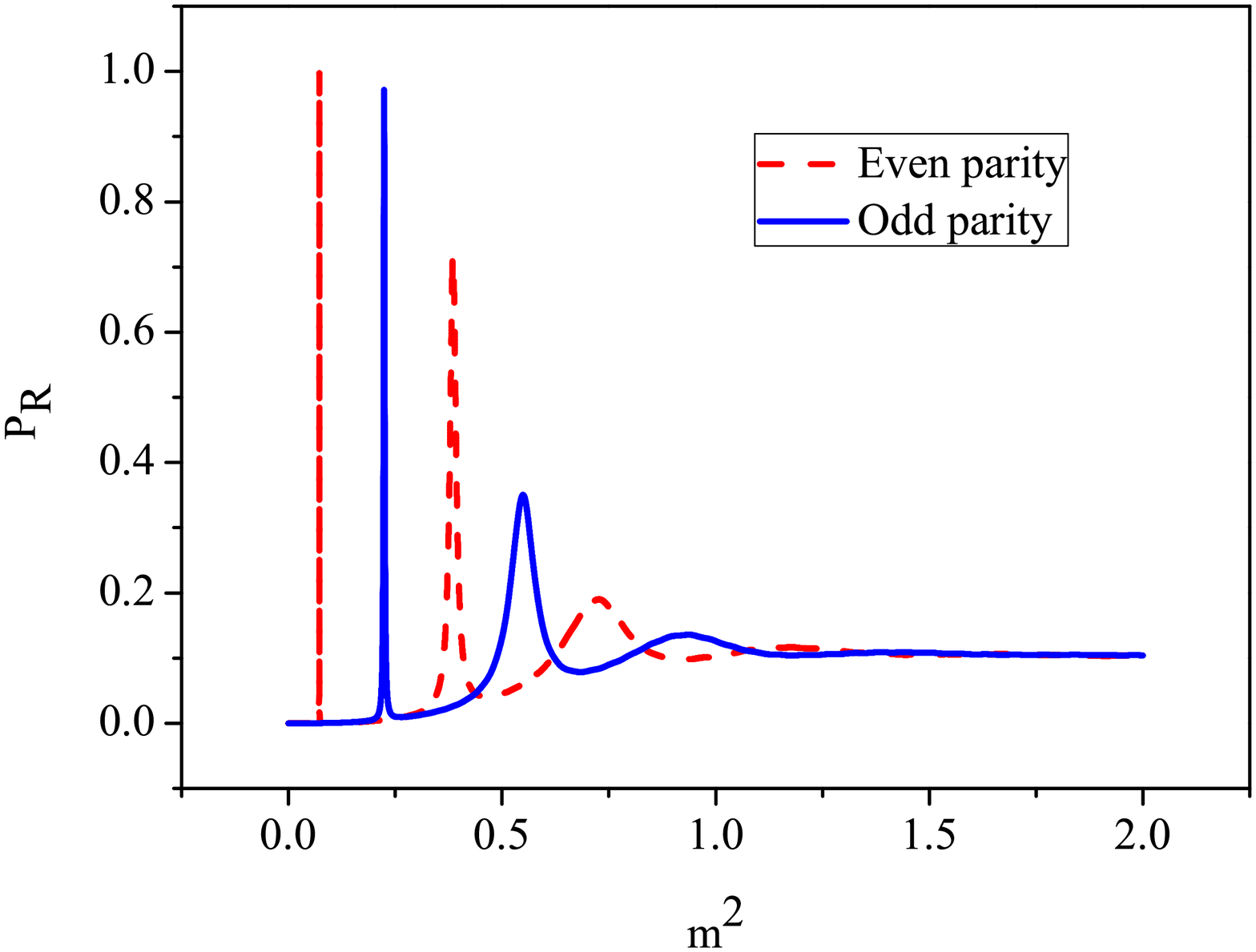}}
\caption{The probability $P_{\text{L},\text{R}}$ (as a function of $m^{2}$)
 for finding massive KK modes of left- and right-chirality
 fermions with mass $m^{2}$ around the brane location
 for $F(\phi)=\phi^{3}$.
 The solid lines and dashed lines are plotted for the odd-parity
 and even-parity massive fermions, respectively. The parameters are set to $z_{\text{max}}=120$, $V_0=0.05$, and $\eta=0.05$.}
 \label{figPunityk=13}
\end{figure*}

For the simplest generalized Yukawa coupling type $F(\phi)=\phi^{k}$, $k=1, 3$, when we choose a $V_0$, some resonant states would be obtained. When the eigenvalues deviate from the mass eigenvalues of the resonances, the resonant states will become unclear and even disappear. The graphics of massive resonant wave functions, potential functions, and off-resonant wave functions are shown in Fig.~\ref{figunifyk=13}. From the figure, we can obviously see the impacts of the potential functions $V_{\text{L}}(z)$ and $V_{\text{R}}(z)$ on the left- and right-handed massive KK modes with even parity or odd parity. From the comparison between the
massive resonant wave functions and massive off-resonant ones, we find that
the resonances occur only at some particular mass eigenvalues, e.g., $m^2=0.072$ (even), $0.072$ (odd), etc. for $k=3$, $\eta=0.05$, and $V_0=0.05$. These mass eigenvalues show that they follow the Kaluza-Klein parity-chirality decompositions of the massive fermion resonances obtained~\cite{Liu:09d}:
\begin{subequations}\label{the general chiral decomposition explicit}
\begin{eqnarray}
\Psi(x,z)&=&e^{-2A}\sum_{n}\left[\psi_{\text{L}n}(x)f^{(\text{E})}_{\text{L}n}(z)
+\psi_{\text{R}n}(x)f^{(\text{O})}_{\text{R}n}(z)\right]\,,\nonumber\\
\label{the general chiral decomposition explicit a}\\
\Psi(x,z)&=&e^{-2A}\sum_{n}\left[\psi_{\text{L}n}(x)f^{(\text{O})}_{\text{L}n}(z)
+\psi_{\text{R}n}(x)f^{(\text{E})}_{\text{R}n}(z)\right]\,,\nonumber\\
\label{the general chiral decomposition explicit b}
\end{eqnarray}
\end{subequations}
where
the superscripts ``$\text{E}$'' and ``$\text{O}$'' stand for even parity and odd parity,
respectively.
As we know, this demonstrates that Dirac fermions can be composed
of the left-handed fermions with odd parity and the right-handed ones
with even parity, and vice versa~\cite{Liu:09b,Liu:09d}.
Specifically, the parities and chiralities of the resonant massive KK
modes are conserved for left- and right-handed resonant massive fermions.

Besides, we note that the potential of left-handed fermions for $k=3$ becomes a double-well potential as discussed in Ref.~\cite{Liu:09b,Liu:09d}, and this can be seen from
Fig.~\ref{figunifyb}. After simple calculations, we find
that the potential of left-handed fermions is always a double-well one for $F(\phi)=\phi^{k}$ with odd $k\geq3$, whereas
the potential of right-handed fermions is a single-well potential all the
time. However, there exist fermion resonances for both kinds of potential wells.

According to the knowledge of quantum mechanics, we know that, when a microscopic particle encounters a potential well, the microscopic particle
will stay in the potential well with a limited lifetime. Next, we will carefully
study the resonant processes. Since the Schr\"{o}dinger equations~(\ref{Scheq}) can
be recast into the form of $\mathcal {O}^{\dag}_{\text{L},\text{R}}\mathcal
{O}_{\text{L},\text{R}}f_{\text{L},\text{R}}(z)=m^{2}f_{\text{L},\text{R}}(z)$, the probability for finding the massive KK
modes around the brane location along extra dimension is $|f_{\text{L},\text{R}}(z)|^{2}$.
From the previous section, we know that the energy density of the thick brane concentrates in a finite range along the direction of the extra dimension. Strictly speaking, we can interpret $|f_{\text{L},\text{R}}(0)|^{2}$ as the probability for
finding the massive KK modes at the center of the brane. However, for the massive
KK modes with odd parity, the wave functions at the center of the brane are
$f^{(\text{O})}_{\text{L},\text{R}}(0)=0$. It means that the probability is poorly defined, and we cannot understand the resonances very well. Therefore,
without loss of generality, we define a relative probability to express the
probability of massive KK modes with even and odd parities around the brane
location. The relative probability is defined as~\cite{Liu:09b}
\begin{eqnarray}
P_{\text{L},\text{R}}\left(m^2\right)=\frac{\int^{z_b}_{-z_b}|f_{\text{L},\text{R}}(z)|^{2}dz}{\int^{z_{\text{max}}}_{-z_{\text{max}}}|f_{\text{L},\text{R}}(z)|^{2}dz}\,,
\label{relative probability}
\end{eqnarray}
where the relationship between $z_b$ and $z_{\text{max}}$ is chosen as $z_{\text{max}}=10z_b$, so the
probability for the plane wave modes with the eigenvalue $m^{2}$ is $1/10$. When we set the same parameters $V_0=0.05$ and $\eta=0.05$ as in Fig.~\ref{figunifyk=13}, for $F(\phi)=\phi^{k}$ with $k=3$, the probabilities $P_{\text{L},\text{R}}$ (as a function of $m^{2}$) for finding massive KK modes of
left- and right-chirality fermions around the brane location are
depicted in Fig.~\ref{figPunityk=13}. We find that the probability is maximized at some mass eigenvalue (a series of resonant peaks) and in the figure the plateau corresponds to $z_b/z_{\text{max}}=0.1$. The number of resonant peaks for left-handed KK modes is equal to that of the ones for right-handed KK modes.
In a sense, this shows that a massive Dirac fermion can be composed of the left- and right-handed massive KK modes.

Furthermore, in the numerical calculations, we find that the magnitude of the
probability for finding a fermion around the brane is related to the step size in
the numerical experiments. Accordingly, in order to analyze the fermion resonances, the
step size of the coordinate used in the numerical calculations must be small enough to reflect the true face of resonances. In the subsequent discussions, we take a small enough step size to study the probability of the resonant states. In this case, more detailed graphs of resonant probabilities with respect to the resonant mass $m$ can be obtained. These
resonant peaks correspond to the massive resonant KK modes with even and odd parity obtained in Fig.~\ref{figunifyk=13}. We can clearly see that more notable details (such as the maximum $P_{\text{max}}$ of resonant probabilities) of resonant peaks are shown compared with the resonant peaks in Fig.~\ref{figPunityk=13}. At the same time, from Fig.~\ref{figPunityk=13}, we can see that when the mass eigenvalue increases, the resonant peaks become thicker and thicker. On the other hand, the resonant peaks with relatively smaller mass eigenvalues are narrower, and the resonant peak with the smallest mass eigenvalue (the first peak) is the thinnest one. Later, we will know that the narrowest resonant peak has the maximum resonant lifetime.

\begin{table}[h]
\begin{center}
\caption{The eigenvalue $m^2$, mass $m$, width $\Gamma$, lifetime
$\tau$, and maximum probability $P_{\text{max}}$ for resonances of left and
right chiral fermions with odd-parity and even-parity solutions for
 $F(\phi)=\phi^{3}$. $\mathcal{C}$ and
$\mathcal{P}$ stand for chirality and parity, respectively.
$\mathcal{L}$ and $\mathcal{R}$ are short for left handed and
right handed, respectively. The parameters are set to $z_{\text{max}}=120$, $V_0=0.05$, and $\eta=0.05$.} \label{TableSpectrak}
\begin{ruledtabular}
\begin{tabular}{cccccccc}
 $k$ & $\mathcal {C}$  & $\mathcal {P}$ &  $m^{2}_{n}$ & $m_{n}$ & $\Gamma$& $\tau$ &  $P_{\text{max}}$   \\
\hline

 &  & Odd   &   0.072      & 0.269  & 0.000~3   & 3~505.3  & 0.999  \\ 
 &      & Even    & 0.224     & 0.474    &  0.003  & 343.0  &  0.971  \\ 
 & \raisebox{1.3ex}[0pt]{$\mathcal{L}$ }   & Odd  & 0.386 & 0.621 & 0.013 & 78.97 & 0.731 \\
  &     & Even    & 0.551      & 0.742     & 0.043    & 22.99  &  0.362  \\ 
 \cline{2-8}\cline{2-8}
  \raisebox{1.3ex}[0pt]{3}&    & Even  & 0.072 & 0.269 & 0.000~3 & 3~422.2 & 0.998
  \\ 
  &      & Odd   &  0.224     & 0.473   & 0.003  & 324.0  &  0.972 \\ 
  &  \raisebox{1.3ex}[0pt]{$\mathcal{R}$ }   &Even    & 0.385  & 0.621      & 0.014    & 71.21   & 0.716  \\ 
    &   & Odd   & 0.549   & 0.741   & 0.051  & 19.63  & 0.350 \\ 

\end{tabular}

\end{ruledtabular}
\end{center}
\end{table}

To further analyze the lifetime of a fermion resonance, first, we define the
width $\Gamma=\Delta m$ of a resonant state as the width at the
half maximum of a resonant peak~\cite{Gregory:00}. In this case, a massive fermion will
disappear into the fifth dimension after staying on the brane for some time $\tau$ $\thicksim$ $\Gamma^{-1}$. Thus, $\tau$ is called the lifetime of a fermion resonance mentioned above. After numerical calculations, we can get a lifetime from each peak of
the fermion resonance. In Table~\ref{TableSpectrak}, we
list the eigenvalue $m^2$, mass $m$, width $\Gamma$, lifetime $\tau$, and maximum probability $P_{\text{max}}$ for resonances of left and right chiral fermions with odd-parity and even-parity solutions for $F(\phi)=\phi^{3}$. It is interesting that the potential wells of the massive chiral fermions are not deep enough for $k=1$, $V_0=0.05$, and $\eta=0.05$, consequently, there is no resonant state in this situation. However, if we choose a larger $\eta$, the height of the potential of the massive chiral fermions will become larger, then a series of massive fermion resonant states will appear.

Note that the data in Table~\ref{TableSpectrak} only reflect the information of the fermion
resonances for $k=3$. After further calculations, we find that, for a given set of parameters,
the number of resonant states increases with $k$. The resonant mass
eigenvalue $m$ of a left-handed fermion with odd parity is equal to that of a
right-handed fermion with even parity. On the other hand, the resonant mass
eigenvalue $m$ of a right-handed fermion with odd
parity is also equal to that of a left-handed fermion with even parity. This means that,
in this case, the massive KK modes satisfy the equations of the Kaluza-Klein
parity-chirality decompositions~(\ref{the general chiral decomposition explicit}). In
other words, a massive Dirac fermion can be composed of the left- and right-handed
massive KK modes, as mentioned above. Indeed, this result can be understood in the context of supersymmetric quantum mechanics. The Hamiltonians in (\ref{HamiltonianLeft}) and (\ref{HamiltonianRight}) can be factorized as $H_\text{L}=A^\dag A$ and $H_\text{R}=A A^\dag$. Hence $H_\text{L}$ and $H_\text{R}$ are actually conjugated supersymmetric partner Hamiltonians with the superpartner potentials $V_\text{L}(z)$ and $V_\text{R}(z)$. This leads to the correspondence between the spectra of left- and right-handed fermions except a normalized zero mode of $f_{\text{L}0}(z)$, i.e. the spectra of $H_\text{L}$ and $H_\text{R}$ are degenerate. The underlying reason for the degeneracy of the spectra of $H_\text{L}$ and $H_\text{R}$ can be understood from the properties of the supersymmetric algebra\cite{Cooper}.

\begin{figure}
\includegraphics[width=0.4\textwidth]{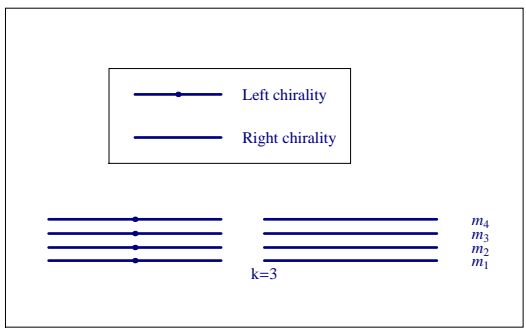}
\caption{The mass spectrum of resonances for
$F(\phi)=\phi^{3}$. The parameters are set to
$z_{\text{max}}=120$, $V_0=0.05$, and $\eta=0.05$.}
\label{massspetra_k}
\end{figure}

For the lifetimes of the massive resonant states, there exists a
following rule. For a relatively smaller resonant mass
eigenvalue $m$, the resonant state has a relatively larger lifetime $\tau$. Actually, this rule can be seen obviously from the profiles of the peaks of the resonant probabilities. From Fig.~\ref{figPunityk=13}, we can clearly see that, when the magnitude of a mass eigenvalue increases, the resonant peak would become thicker. The resonant peaks with
relatively huger mass eigenvalues would be broader, and the
resonant peak with the biggest mass eigenvalue (the last peak)
is the thickest one. When the resonant peaks become wide enough, the
massive KK modes will not stay on the brane, and the time scale
$\tau\rightarrow0$. Therefore, the massive KK modes will tunnel
through the brane at a moment, consequently, they will become free both in
the bulk and on the brane. On the contrary, for the
resonant state with the smallest mass eigenvalue, it will have a very long time to stay on the brane. The massive KK modes with a huge lifetime $\tau$ will gradually pass through the
brane, and slowly fade away into the bulk. The extreme situation is that, when $\tau$ tends to infinity, i.e., $\tau\rightarrow\infty$, the resonant state will become a bound
state.

In addition, we can plot the mass spectrum of the massive resonant KK modes. For $F(\phi)=\phi^{k}$ with $k=1, 3, 5,\ldots$, the mass
spectrum of the massive fermion resonances is depicted in Fig.~\ref{massspetra_k} for $k=3$.
We can see that the number of the resonant states of the massive
fermions with left- and right-handed chiralities increases with $k$. However, in the later discussions, we will find that the situation is slightly different. The resonant states here are the ones whose resonant mass eigenvalues are below the maxima $V_{\text{L,R,max}}$ of the potentials (note that $V_{\text{L,max}}$ and $V_{\text{R,max}}$ are the same magnitude). In general, the number of the right-handed resonant states of the massive KK modes is equal to that of the ones with left-handed chirality.

\subsubsection{Case II: $V_0$ as a variable}\label{secV0}

In this subsection, we mainly discuss the impacts of the brane parameter $V_0$ (it decides the thickness of the brane by the relation $d={5.074~5}/{\sqrt{V_0}}$) on the resonant states of the massive KK fermions, especially on the resonant probabilities, resonant masses, and the number of resonant states. We find that the impacts of $V_0$ and $k$ on the resonances are different.

For given $\eta$, $k$, and various values of $V_0$, we study the resonances of massive KK modes. As investigated in the previous subsection, using the Numerov method, we numerically solve the Schr\"{o}dinger
equations~(\ref{Scheq}) with the potential functions~(\ref{VzCaseI}). Following the
same routine used above, for $F(\phi)=\phi$, the eigenvalue $m^2$,
mass $m$, width $\Gamma$, lifetime $\tau$, and maximal probability $P_{\text{max}}$ are
listed in Table~\ref{TableSpectraV0} for massive resonances of left and
right chiral fermions with odd and even parities.

\begin{table}[h]
\begin{center}
\caption{The eigenvalues $m^2$, mass $m$, width $\Gamma$, lifetime
$\tau$, and maximum probability $P_{\text{max}}$ for resonances of left and
right chiral fermions with odd parity and even parity for
$F(\phi)=\phi$. $\mathcal{C}$ and
$\mathcal{P}$ stand for chirality and parity, respectively.
$\mathcal{L}$ and $\mathcal{R}$ are short for left- handed and
right handed, respectively. The parameters are set to $\eta=1$, $z_{\text{max}}=30$ for $V_0=0.1$, and $z_{\text{max}}=60$ for $V_0=0.5$.}
\label{TableSpectraV0}
\begin{ruledtabular}
\begin{tabular}{cccccccc}
 $V_0$ & $\mathcal {C}$  &  $\mathcal {P}$ & $m^{2}_{n}$ & $m_{n}$ & $\Gamma$& $\tau$ &  $P_{\text{max}}$   \\
 \hline

 &   &  odd & 0.815      & 0.902   & 0.004   & 266.1  & 0.987 \\
 & \raisebox{1.3ex}[0pt]{$\mathcal{L}$}    &  even    & 1.417        & 1.191      & 0.047  & 21.34 &  0.548   \\  \cline{2-8}
  \raisebox{1.3ex}[0pt]{0.1}&      &  even & 0.814 & 0.902 & 0.004 & 261.6 & 0.989
  \\
  & \raisebox{1.3ex}[0pt]{$\mathcal{R}$}    &  odd  & 1.416    & 1.190     & 0.049   & 20.44 & 0.562 \\ \hline

 &  $\mathcal{L}$    &  odd & 1.547      & 1.244    & 0.152   & 6.569 & 0.408 \\
  \raisebox{1.3ex}[0pt]{0.5}&   $\mathcal{R}$    &  even & 1.522 & 1.234 & 0.166 & 6.021 & 0.418 \\

\end{tabular}
\end{ruledtabular}
\end{center}
\end{table}

From Table~\ref{TableSpectraV0}, we can clearly see that
all the values of the eigenvalue $m^2$, mass $m$, width $\Gamma$,
lifetime $\tau$, and maximal probability $P_{\text{max}}$ follow the same
rules obtained in the previous subsection. However, when
$V_0$ becomes larger, the results are quite different from those
with smaller $V_0$. In this case, we find that the maximum values $V_{\text{L},\text{max}}$ and $V_{\text{R},\text{max}}$ of the potential functions of the left- and right-handed KK modes are clearly the same magnitude. The potential for the smaller $V_0$ is deeper than that of ones for bigger $V_0$, while the height of potential remains unchanged for various values of $V_0$. Consequently, as $V_0$ increases, the number of resonant states decreases.

We find that the probability of the first peak of resonant states with right-handed chirality is the maximum, and it is equal to that of the one with left-handed chirality. For both chiralities, as the resonant mass eigenvalue increases, the resonant probability decreases. The resonant probabilities of left- and right-handed massive KK fermions have a good match at the same resonant mass eigenvalue. Accordingly, there is a one-to-one correspondence between the peaks of left- and right-handed fermions at the same resonant mass eigenvalue. As already discussed, the mass spectrum can be plotted from the data in Table~\ref{TableSpectraV0}. Until now, we have seen that there exists a good match of the probabilities $P_{\text{L}}$ and $P_{\text{R}}$ for all of the excited states and its origin is the consistency of the behavior of the potentials $V_{\text{L}}$ and $V_{\text{R}}$. We can see that, for each $n$, which labels the $n$th resonance, within some given numerical error, the resonant mass $m_{n}$ of the resonant KK modes with left-handed chirality is equal to that of ones with right-handed chirality. In other words, a massive Dirac fermion could also be composed of the left- and right-handed massive KK modes.

In a word, in this section we mainly investigate the resonant problem of
the massive chiral KK modes with odd and even parities. For various
values of $k$ and $V_0$, we obtain the eigenvalue $m^2$, mass $m$, width $\Gamma$, lifetime $\tau$, and maximum probability $P_{\text{max}}$ for the resonances of left and right chiral fermions with odd and even parities for $F(\phi)=\phi^{k}$ with odd $k=1,3$. The results can be extended to $F(\phi)=\phi^{k}$ with odd $k=5,7,\cdots$. Besides, we obtained the graphs of the mass spectra of the massive KK resonant states. For the two cases, as discussed in Refs.~\cite{Liu:09b,Liu:09d}, the resonant states of the massive KK modes satisfy the Kaluza-Klein parity-chirality decompositions. As the parameter $V_0$ becomes larger, the depth of the potential wells $V_\text{L}$ and $V_\text{R}$ becomes shallower, while the height of the potentials remains unchanged. Consequently, as $V_0$ increases, the number of resonant states decreases. In other words, when the thickness of the brane becomes thicker, the number of resonant states increases. The resonant probabilities $P_{\text{L}}(m_n)$ and $P_{\text{R}}(m_n)$ of the resonant massive KK modes have a good match at some resonant mass eigenvalue for all of the excited states.

\section{Discussions and conclusions}
\label{secConclusion}
In this paper, we have investigated the fermion resonances in the
background of a single scalar field generated thick brane with a piecewise warp factor. Using the warp factor inspired by the one in Ref.~\cite{0801.0801}, we obtained the kinklike
background scalar field without derivative jumps under three junction conditions. The distance $d$ between the two boundaries, which is just the thickness of brane, can be obtained in terms of one free parameter $V_0$. Next, we introduced the interaction between a massless bulk fermion and the background real scalar field by means of a general Yukawa coupling $\eta\overline{\Psi}F(\phi)\Psi$ in five-dimensional
space-time, where $F(\phi)=\phi^{k}$ with odd $k=1,3,5,\ldots$.

When the parameter $k$ is treated as a variable, we found
that the coupling between the scalar and fermion influences the resonances. The number of resonant states of the massive KK fermions with the left- and right-handed chiralities increases with $k$. Here, the number of resonant states is defined as the number of those resonant states with eigenvalue $m^{2}$ lower than the highest point of the potential. We found that the number of the resonant KK
modes with left-handed chirality is equal to that of the ones with right-handed chirality. This
means that the fermion resonant states of massive KK modes satisfy the Kaluza-Klein parity-chirality decompositions~\cite{Liu:09d}. For every $k=\{1,3,5,\ldots\}$, the resonant KK fermions with
smaller mass eigenvalues have larger lifetimes than those with bigger mass eigenvalues. On the other hand, we can see that the wider resonant peaks have smaller lifetimes. Furthermore, the first resonant peak is the narrowest one, and it has the longest life expectancy. Accordingly, the KK fermion resonances with smaller mass eigenvalues can stay on the thick brane for a longer time. In other words, these quasibound states (resonant states) are more stable. Conversely, the KK fermion resonances with huger mass eigenvalues can only stay on the brane for a little time, i.e., the resonances become too unstable and cease to appear~\cite{Almeida:0901}.

When the brane parameter $V_0$ is treated as a variable, we found that the thickness of the brane also influences the fermion resonances. The number of the peaks of resonant probability of the massive KK fermions with left- and right-handed chiralities decreases with $V_0$, namely, the number of resonant states decreases with $V_0$. When $V_0$ varies, the maximum values $V_{\text{L},\text{max}}$ and $V_{\text{R},\text{max}}$ of the potential functions of the left- and right-handed KK modes are the same magnitude. Accordingly, the probabilities of the resonant KK modes of left-handed fermions are equal to that of the ones of right-handed fermions. Further, the masses and lifetimes of left and right chiral resonances are almost the same, which demonstrates that it
is possible to compose a massive Dirac fermion from the left and right chiral resonances. In this case, the resonant probabilities of the left- and right-handed resonant KK modes with the same mass are the same for all the massive resonant KK modes. There is a good match for the resonant probabilities of the fermion resonances with left and right chiralities.

Indeed, if a resonance is supposed to represent a metastable four-dimensional massive fermion field, it must contain both left and right parts, which are mixed by the four-dimensional Dirac equations [see Eqs.~(\ref{spinorCoupledEqs1}) and (\ref{spinorCoupledEqs2})]. We can see that there is no mismatch for peaks of the probabilities $P_{\text{L}}$ and $P_{\text{R}}$ (as a function of $m^{2}$) for finding massive KK modes of left- and right-chirality fermions. The numbers of peaks of $P_{\text{L}}$ and $P_{\text{R}}$ are equal to each other. So the resonant masses of the left- and right-handed fermions must be equal to each other. So our computations are consistent with the coupled four-dimensional Dirac equations. Actually, the definition of the number of resonant states in Ref~\cite{Liu:09d} can be used because there is a good match for the maximum values $V_{\text{L},\text{max}}$ and $V_{\text{R},\text{max}}$ of the potential functions of the left- and right-handed KK modes. From the definition, we can see that only the resonant states with eigenvalue $m^{2}$ lower than the highest point of the potential are considered as resonant states because they have obvious resonant peaks, namely, they have relatively larger resonant probabilities. In other words, the resonant states with unobvious resonant peaks were neglected. So the resonant states with larger mass eigenvalues (the ones with quite small resonant probabilities) were not taken into account. The prior condition is that the maxima of the potential functions $V_{\text{L}}$ and $V_{\text{R}}$ are equal to each other, or at least they have the same magnitude. Actually, the numbers of resonant states can be considered as numbers of peaks of $P_{\text{L}}$ or $P_{\text{R}}$. Although the resonant probabilities (whose resonant masses is larger than the threshold value of $V_{\text{R},\text{max}}$) are quite small, we should take them into account. Thus, from the physical point of view, all the resonant peaks describe the resonant states. Essentially, the resonant behavior coincides with the qualitative change of the energy density of the brane.

With respect to resonances for chiral fermions on the thick brane with a piecewise warp factor, we can resume our conclusions with the following points: (i) The thick brane with a piecewise warp factor was derived under three junction conditions which ensure the correct Israel junction conditions. The kinklike scalar field without derivative jumps can  be given. Especially, the continuous scalar potential and energy density were also obtained. (ii) The thick brane without internal structure also favors the appearance of resonant states for both left- and right-handed fermions. When the brane becomes thicker ($V_0$ becomes smaller), the number of resonant states increases, and vice versa. (iii) The scalar-fermion coupling influences the resonant behaviors of the KK fermions. The resonant properties (such as resonant probability, lifetime, mass, and number) of the left-handed fermions are the same as the ones of the corresponding right-handed fermions. (iv) Strictly speaking, all of the resonant peaks describe the resonant states. The definition of the number of resonances given in Ref.~\cite{Liu:09d} should be redefined. Indeed, the number of resonant states can be considered as number of resonant peaks. The fermion resonant states of the massive KK modes satisfy the Kaluza-Klein parity-chirality decompositions. (v) It is worth noting that many systems with boundaries cannot be investigated in the thick brane scenario, instead the thin branes as the boundary terms should be added into the action. Such systems can be studied with the method in Ref.~\cite{DeWolfe:00a}. The properties of the fermion resonances in thick brane models and the problems in the brane with boundaries are rich and interesting. We leave these issues for future works.

\section{Acknowledgements}

Helpful comments from the anonymous referee, M. Cveti\v{c}, A. Herrera-Aguilar, and C. Kokorelis are gratefully acknowledged. Beneficial discussions with Ke Yang are also highly appreciated. This work was jointly supported by the Program for New Century Excellent Talents in University, the Huo Ying-Dong Education Foundation of Chinese Ministry of Education (No. 121106), the National Natural Science Foundation of China (No. 11075065), the Doctoral Program Foundation of Institutions of Higher Education of China (No.
20090211110028), the Key Project of Chinese Ministry of Education
(No. 109153), and the Natural Science Foundation of Gansu
Province, China (No. 096RJZA055).

\end{document}